\documentclass{ws-ijmpa}

\begin{document}

\markboth{M. Giannotti}
{Mirror World and Axion: Relaxing Cosmological Bounds}

%
\catchline{}{}{}{}{}
%

\title{MIRROR WORLD AND AXION: RELAXING COSMOLOGICAL BOUNDS}

\author{\footnotesize MAURIZIO GIANNOTTI}

\address{Dipartimento di Fisica and Sezione INFN di Ferrara,\\
        via Paradiso 12, I-44100 Ferrara, Italy\\ giannotti@fe.infn.it}

\maketitle

\pub{Received (Day Month Year)}{Revised (Day Month Year)}

\begin{abstract}
The cosmological (upper) limit on the Peccei-Quinn constant,
related to the primordial oscillations of the axion field,
can be relaxed for a mirror axion model. The simple reason is
that the mirror world is colder and so the behavior of 
the axion temperature-dependent mass is dominated by the 
contribution from the mirror sector. So the coherent oscillations 
start earlier and correspondingly the axion mass density 
$\Omega_a h^2$ is reduced. 

\end{abstract}

\bigskip\bigskip

\noindent
The CP violating term
$\mathcal{L}_{\Theta}=\Theta(\alpha_s/8 \pi)G_{\mu\nu}\tilde G^{\mu\nu}$,
in the QCD Lagrangian,
leads to observable effects (e.g. non zero neutron electric dipole moment),
experimentally not observed.
This fact is referred to as the strong CP problem 
(for a general reference see Ref.~\refcite{kim}).

In the Peccei-Quinn
(PQ) mechanism,\cite{PQ} which is the most appealing solution of this problem,
the parameter $\Theta$ becomes a
dynamical field, the axion $a=(f_{\rm PQ}/N)\Theta$, 
whose potential is minimized for the CP even configuration $\langle a \rangle =0$.
The axion emerges as the
pseudo-Goldstone mode of a spontaneously broken global axial
symmetry $U(1)_{\rm PQ}$. Here $f_{\rm PQ}$ is a constant, with
dimension of energy, called the PQ constant, and $N$ stands for
the color anomaly of $U(1)_{\rm PQ}$
current.\cite{kim}
In the following, we will usually refer
to the constant $f_a=f_{\rm PQ}/N$, which characterizes the axion phenomenology.
Astrophysical considerations exclude all the value of $f_{a}$ 
up to $10^{10}$ GeV.\cite{kim,raffelt} On the other hand, the cosmological limit,
related to the primordial oscillations of the axion field,
demands the upper bound 
$f_a<10^{12}$ GeV.\cite{kim,raffelt} 
Let us briefly review the origin of this last limit.

In most axion models, the PQ symmetry breaking occurs when a complex scalar 
field, $\phi$, which carries PQ charge, acquires a vacuum expectation value (VEV)
$\langle \phi \rangle \sim f_{\rm PQ}$.
This occurs as the temperature of the universe cools down below the value of 
the PQ temperature. As said, the axion is identified with the angular degree of
freedom $a=f_a\Theta$. Today $\Theta$ is settled in the CP-conserving 
minimum $\Theta =0$. However, after the PQ symmetry breaking, 
as the temperature is of the order of the PQ scale, the axion is massless 
and the initial value of the phase $\Theta$ is chosen stochastically.
We indicate this initial value with $\Theta_i$.

When the universe is cold enough for the QCD phase transition, instanton effects
curve the potential so that the axion acquires a non zero mass 
$m_a(T)=m Q(T)$, where $m$ is the zero temperature limit:\cite{kolb}
\begin{equation}\label{a-m}
	m\simeq 0.5~ \frac{f_{\pi} m_{\pi}}{f_a}\simeq 0.62 ~{\rm eV} \frac{10^7 {\rm GeV}}{f_{a}}~.
\end{equation}
\noindent
The function $Q(T)$ for temperature $T\gg \Lambda$
($\Lambda \sim 200$ MeV is the QCD scale) 
was calculated in Ref.~\refcite{GPY} and
can be approximated as:\cite{Turner}
\begin{equation}\label{f}
Q(T)= A\left(\frac{\Lambda}{T}\right)^{b}~,
\end{equation}
\noindent
where $A=0.1\times 10^{\pm 0.5}$ and $b=3.7\pm 0.1$.
This power law is valid when $T>\Lambda$ while for
$T \ll \Lambda $ we have $Q=1$ and the mass becomes independent of the
temperature. We use this relation until $Q(T)<1$, that is until $T > T_c =
A^{1/b}\Lambda \simeq 100$ MeV, and assume that $Q(T)=1$ for $T< T_c$.

Then $\Theta$ starts to roll down, slowed by the
Hubble expansion, following the equation: 
\begin{equation}\label{Theta}
\ddot{\Theta} + 3H(T) \dot{\Theta} + m_a^2(T)\Theta = 0~,
\end{equation}
\noindent
where $H(T)=1.66 g_\ast^{1/2}T^2/M_{Pl}$
is the Hubble expansion rate.
When the curvature term dominates on the (Hubble) friction term,
$\Theta$ begins to oscillate with the frequency $m_a(T)$. This happens
at the temperature value $T=T_i$ defined by the equation
$m_a(T_i)=3 H(T_i)
\approx 5 g_{\ast i}^{1/2}T_i^2/M_{Pl}$,
where $M_{Pl}=1.22\times 10^{19}$ GeV is the Plank mass 
and $g_{\ast i}=g_\ast(T_i)$ is the effective number of the 
particle degrees of freedom at $T=T_i$.\footnote{For the 
standard particle content, 
$g_{\ast}(T) = 110.75$ for $T> 100$ GeV, 
86.25 for 100 GeV $> T >$ 5 GeV,
75.75 for 5 GeV $> T >$ 2 GeV,
61.75 for 2 GeV $> T > \Lambda$,
17.25 for $\Lambda > T >$ 100 MeV,
10.75 for 100 MeV $> T >$ 1 MeV.}

For temperatures $T < T_i$, when the coherent oscillations commence,
eq. (\ref{Theta}) can be considerably simplified, and leads to the quite
simple result $n_a R^3 = $ const, where $n_a(T) = \rho_a(T)/m_a(T)$
is the axion number density. 
In other words, the number of axions in a comoving volume 
remains in fact constant during the universe expansion.  
We further assume that no entropy production takes place 
after the moment when the axion field begins to oscillate. 
Then the axion number density
to the entropy density ratio, $\eta_a=n_a(T)/s(T)$,
remains constant at $T<T_i$, and thus the present energy density of axions is 
$\rho_a^0=\eta_a s_0m$, 
where $s_0=(2\pi^2/45) g_0^{s}T_0^3$ is the present entropy
density, $g_0^{s}=3.91$ and $T_0=2.725$ K is the 
CMB temperature. At the initial moment $t_i$, the axion 
number density is $n_a(T_i) \simeq \rho_a(T_i)/m_a(T_i) \simeq 
\frac12 m_a(T_i) \Theta_i^2 f_a^2$. Comparing to the critical density 
$\rho_{\rm cr} = 8.1h^2 \cdot 10^{-47}$ GeV$^4$, 
we obtain:
\begin{equation}\label{Om-T1}
\Omega_a h^2 = 0.77 \, 
\frac{\Theta_i^2 f_{12} } {g_{\ast i}^{1/2} T_{i1} } ~, 
\end{equation}
\noindent 
where $T_{i1} = (T_i/1\, {\rm GeV})$
and $f_{12}=f_a/10^{12}$ GeV.

Substituting (\ref{f}) in the expression $m_a(T_i)=3 H(T_i)
\approx 5 g_{\ast i}^{1/2}T_i^2/M_{Pl}$,
which defines the temperature $T_i$, we obtain:
\begin{equation}\label{T1}
T_i =  \Lambda
\left(\frac{A m M_{Pl}}{5 g_{\ast i}^{1/2}\Lambda^2 } 
\right)^{\frac{1}{b+2}} \simeq 
f_{12}^{-0.175}\Lambda_{200}^{0.65}\times (0.9\pm 0.2) ~ {\rm GeV}~,  
\end{equation}
\noindent
where $\Lambda_{200} = (\Lambda/200\,{\rm MeV})$ and the 
uncertainties of about 20 percent are related to uncertainties 
in parameters $A$ and $b$.
This estimation very weakly depends on $f_{12}$ and thus 
for the Peccei-Quinn scales of the cosmological relevance, 
starting from the astrophysical bound $f_a \geq 10^{10}$ GeV 
to $f_a$ order GUT scale $10^{15-16}$ GeV, $T_i$ changes 
within interval 2 GeV -- 200 MeV  
and hence $g_{\ast i}^{1/2} \approx 8$. 
This estimation is valid, however, until 
$T_i > T_c = A^{1/b}\Lambda \simeq 100$ MeV.\footnote{For $T < T_c$ 
instead we have $m_a(T)=m$, 
$g_\ast(T)^{1/2}\approx 4$ and hence 
$T_i \simeq f_{12}^{-1/2} \times 60$ GeV. 
This condition holds for 
$f_a > 3.2\cdot 10^{17}\Lambda_{200}^{-2}$ GeV. }

Therefore, we obtain: 
\begin{equation}\label{Omegah2-st}
\Omega_a h^2 \simeq 0.10 \, 
\Lambda_{200}^{-0.65} \Theta_i^2 f_{12}^{1.175}~.
\end{equation} 
\noindent
Demanding $\Omega_a h^2<0.15$, the natural choice $\Theta_i\sim 1$, 
leads to $f_{12}\mathrel{\mathop  {\hbox{\lower0.5ex\hbox{$\sim$}
\kern-1.1em\lower-0.7ex\hbox{$<$}}}} 1.5$.\footnote{If the axion phase 
transition took place after inflation, 
then for the initial value $\Theta_i$ one implies the {\it rms} 
average from a uniform distribution of initial values from 
$-\pi$ to $\pi$, $\overline{\Theta}_i=\pi/\sqrt3$. Thus, in this case 
we obtain an upper limit on $f_{12}<0.5$.
However, if the PQ symmetry were broken before or during inflation, 
then the value of $\Theta_i$ is the same in the whole Universe within 
the single inflationary patch, and in fact it is an arbitrary 
parameter selected by some random choice.}

It is worth underlining that the only upper limit on
the PQ scale comes from cosmology.
This limit is a few order of magnitude less than the typical 
GUT scale. 
Thus, due to cosmological considerations, it seems quite improbable  
to insert the axion in any GUT model. The obvious question is then: 
how universal is relation (\ref{Omegah2-st})? 
We will show that for a mirror axion model this
can change and, consequently, the cosmological
bound can be relaxed. 


The idea of mirror world was introduced many years ago and 
has been successfully applied in different sectors of 
astrophysics and cosmology.\cite{mirror} 
The mirror world consists of another, parallel, sector of "mirror" 
particles and interactions with the Lagrangian 
completely similar to that of the ordinary particles. 
In other words, it has the same gauge group and coupling constants 
as the ordinary world, 
so that the Lagrangian of the whole theory is invariant with respect 
to the Mirror parity (M-parity) which interchanges the two sectors. 

The possibility to implement the PQ mechanism in the mirror world scenario
is discussed in Ref.~\refcite{mirroraxion}. The general feature is the following:
the total Lagrangian must be of the form $L+L^\prime+\lambda L_{int}$, where 
$L$ represents the ordinary Lagrangian, $L^\prime$ the mirror one and
$L_{int}$ is an interaction term. For $\lambda=0$, the total 
Lagrangian contains two identical $U(1)_{axial}$ symmetries, while the
$L_{int}$ term breaks these in just the usual $U(1)_{\rm PQ}$.
Thus only one axion field results, which is defined in both sectors.

As long as the M-parity is an exact symmetry, the particle physics is 
exactly the same in the two worlds, and so the strong CP problem is 
simultaneously solved in both sectors.  In particular, the axion
couples to both sectors in the same way and their
non-perturbative QCD dynamics produce the same contribution to
the axion effective potential.  This situation does not bring
drastic changes of the axion properties; just the axion zero temperature mass is
increased by a factor of $\sqrt2$ with respect to the standard expression
(\ref{a-m}). On the other hand, on a cosmological ground, the physics is less trivial.
Cosmologically the mirror sector cannot be identical to ours.
If so, the energy density due to the extra (mirror) degrees of freedom would be
much bigger than  what is allowed in the standard cosmology scenario.
The mirror sector must then be colder than ours by a factor 
$x=T^\prime/T< 0.64$.\footnote{More precisely, all the mirror degrees of freedom 
are equivalent to $\Delta N_{\nu}\simeq 6.14$ extra neutrinos, 
in evident contradiction with 
the observation of the Helium abundance predicted by Big Bang Nucleosynthesis
(BBN).
The value $x= 0.64$ corresponds to $\Delta N_{\nu}=1$. For more details 
see Ref.~\refcite{bcv}.
For a complete discussion on the limit on $\Delta N$ allowed by BBN, e.g.
see Ref.~\refcite{fra1} and references therein.}
As a consequence, the energy density of the whole 
universe is essentially the ordinary one,
and the Hubble expansion is completely driven by our sector:
$	H(T)\sim 1.66 g_*^{1/2} T^2/M_{Pl}$,
where $g_*$ counts only the ordinary degrees of freedom.


Let us consider, now, the primordial oscillations of the axion
field for a generic mirror axion model.
For this case, we have $m_a=\sqrt2 m$
in the zero temperature limit. However, at finite temperatures 
the behavior of the axion mass is dominated by the contribution 
from the mirror sector, which has a lower temperature than the 
ordinary one, $T' \ll T$, and thus we have
$m_a(T)\approx m A(\Lambda/xT)^b$. Therefore, the axion mass 
grows faster with the temperature, and the temperature 
$T_i$ at which the axion field starts to oscillate scales 
as $x^{-0.65}$. Correspondingly, the axion mass density 
$\Omega_a h^2$ roughly scales as $\sqrt2 x^{0.65}$. 
E.g. for $x=0.1$, we obtain $\Omega_a h^2$ about 3 times 
smaller than in the estimate (\ref{Omegah2-st}).

However, this situation is valid until the condition 
$T_i(x) > T_c(x)$, where $T_c(x)$ is defined as the temperature
at which $m_a(T)=m$.
This condition roughly translates in $x > 0.01 f_{12}^{1/2}$. 
In this case we obtain:
\begin{equation}\label{temper}
T_i =\sqrt{\frac{m M_{Pl} }{5 g_{\ast i}^{1/2} } } 
\approx \frac{123~{\rm GeV}}{ g_{\ast i}^{1/4} f_{12}^{1/2} }~,
\end{equation}
\noindent
and hence:
\begin{equation}\label{xll1}
\Omega_a h^2 = 2.7\cdot 10^{-3} \Theta_i^2 f_{12}^{3/2}~,
\end{equation}
\noindent 
which, for $\Theta_i\sim 1$, leads to the upper limit $f_{12}<15$.


In conclusion, we have considered here a very simple model for
mirror axion. In fact, in this, the axion properties are 
essentially the same as for the standard invisible axion, and
only the mass is increased by a factor $\sqrt{2}$.
On the other hand, we have shown that its cosmology
is non trivial, because of the different temperature of the mirror
sector. This leads to a weaker cosmological bound on the PQ 
scale. In particular it can be closer to the GUT scale.

\section*{Acknowledgments}

This talk is based on the work in Ref.~\refcite{BGG2}.
I am grateful to the organizers of the Sixth Alexander Friedmann 
International Seminar on Gravitation and Cosmology, for inviting 
me to this interesting conference.

\end{document}